\documentstyle[12pt]{article}
\begin{document}

\begin{center}
{\bf \Large
Elastic Charge Form Factors of $\pi $ and $K$ Mesons
}
\vskip 1cm
{ \large
E.V.Balandina$^1$, A.F.Krutov$^2$ and V.E.Troitsky$^1$
}\\
$^1$ {\it Institute of Nuclear Physics, Moscow State University}\\
$^2$ {\it Samara State University}
\end{center}

\vskip 2cm
\centerline{Abstract}
The elastic charge form factors of the charged $\pi$ and $K$ mesons
are
calculated in modified impulse approximation using
instant form  of  relativistic Hamiltonian dynamics.
Our approach gives
pion and kaon electromagnetic form
factors in the large range of momentum transfer.  The results are in good
agreement with the available data.
Relativistic effects are
large at all values of momentum transfers.

The pion and kaon form factors at large $Q^2$ depend strongly on the
choice of model. The experiments on pion form factor at large momentum
transfer planned at CEBAF will choose between such models.
In the case of kaon such a choosing may be performed only if
supplemented  by accurate measurements of kaon MSR.

\newpage

\rule{0pt}{3cm}
In recent years the interest
has been renewed \cite{BuW94}-\cite{Sch94b} to the study of
electromagnetic form factors of pseudoscalar mesons and,
particularly, of $\pi$ and $K$ mesons.
This fact is due, first of all, to the
experiments planned at CEBAF and concerning the measurement
of pion (E-93-021) and kaon (E-93-018) charge form factors
in the range of momentum transfer  $Q^2 < 3\ $GeV$^2$.
It is possible that these future experiments
will enable to choose between different theoretical
models whose predictions differ rather strongly.

Such a difference of theoretical results seems to be
quite natural, because one encounters a lot of difficulties
while calculating the structure of composite
particles containing light quarks.
The main difficulties are the following two.
First, the importance of relativistic effects
\cite{ChC88}, \cite{Sch94b}
in the whole range of momentum transfer, including
$Q^2 \approx 0$ (the value of mean square radius (MSR))
\cite{KrT93}.
Second, the problem of calculation of the "soft"
and "hard" structure.
The hard part can be calculated from
perturbative QCD. The soft part, which describes the structure at long and
intermediate distances, is modeled in a variety of nonperturbative ways.
The relativistic Hamiltonian dynamics (RHD)
(for a review see \cite{KeP91})
is one of such ways. RHD unifies potential approach to composite systems
and the condition of Poincar\'e-invariance.
This  method
is based on
the direct realization of the Poincar\'e group algebra or, in other words,
of the relativistic invariance condition in the few body Hilbert space.
RHD can be formulated in different ways ( different relativistic dynamics).
At present the light front dynamics is the most popular one and
widely used
to the calculation  of
pion and kaon electromagnetic structure (see for example \cite{ChC88} -
\cite{KiW94}, \cite{Sch94b}).

In this Letter we present a relativistic treatment of the
problem of soft electromagnetic structure in the framework of alternative
form of relativistic dynamics, the so called instant form (IF) of RHD.
Our approach, which provides good
description of the available data for the elastic charge form factors for
the charged $\pi$ and $K$ mesons, will be briefly outlined here.  The
details are partially given in \cite{BaK95} and will be given elsewhere.
The IF of relativistic dynamics, although not widely used, has some
advantages:  first, the calculations can be performed in a straightforward
way; second, this approach is obviously rotationally invariant (this fact
is particularly important for treating spin effects and polarization
effects).  However, the approach encounters some difficulties of general
kind -- the problem of electromagnetic current conservation and of its
relativistic covariance \cite{CoO75} - \cite{Lev95}.

We hope that we have found the way to overcome these difficulties.

Let us consider $\pi$ meson and $K$ meson as
quark ($q$) --- antiquark ($\bar Q$) composite system.
We shall use different quark masses
$M_q$ and  $M_{\bar Q}$ as in $K$ meson.
The results for $\pi$ meson can be obtained if $M_q = M_{\bar Q}$.

The charge form factor for two-quark system can be
obtained from the electromagnetic current matrix element
for composite system
\begin{equation}
<p_c|\,j_\mu\,|{p'}_c>=(p_c+{p'}_c)_\mu\,F_c (Q^2),
\label{pi-par}
\end{equation}
$F_c
(Q^2)$ -- electromagnetic form factor of composite system,
$p_á$ -- 4-mo\-men\-tum of system.

We shall act following the basic assumptions, valid for all forms
of dynamics in RHD \cite{KeP91}.
The RHD
is based on the including of the operator, describing
$q\bar Q$ interaction in the generators of Poincar\'e group while the
commutation relations of Poincar\'e algebra are fulfilled.
One usually
includes $\hat U$ in the mass square operator of the free two particle
system in additive way \cite{KeP91}: $P^2=(p_1+p_2)^2 \>\rightarrow\>{\hat
M}^2_I=P^2+\hat U$. In the case of IF dynamics the Poincar\'e
algebra is conserved if $\hat U$ commutes with the total angular momentum
operator $\hat {\vec J} = ( \hat J_1, \hat J_2, \hat J_3 )$, with the operator
of total 3-momentum $\hat {\vec P}$ and with the operator $\vec \nabla _P$.
The complete set of commuting operators for the two-particle system with
interaction contains now: ${\hat M}^2_I,\>{\hat J}^2,\>\hat J_3,\>\hat
{\vec P}$.
In the case of IF the operators ${\hat J}^2,
\>\hat J_3,\>\hat {\vec P}$ coincide with the
appropriate operators of the two-particle system without interaction and one
can construct the basis in which these three operators are diagonals. While
working in this basis to obtain the wave function one needs to diagonalize
${\hat M}^2_I$.

In RHD the Hilbert space of composite particle states is the tensor
product of single particle Hilbert spaces:
${\cal H}_{q\bar Q} \equiv  {\cal H}_{q} \otimes
{\cal H}_{\bar Q}$
and the state vector in RHD is a superposition of two-particle
states. As a basis in
${\cal H}_{q\bar Q}$ one can choose the following set of vectors:
\begin{eqnarray} |\,\vec p_1\,,m_1;\,\vec p_2\,,m_2\,\!> =
|\,\vec p_1\,,m_1\,\!>\otimes |\, \vec p_1\,,m_2\,>,\nonumber\\ <\,\!\vec
p\,,m\,|\,\vec p\,'\,,
m'\,\!> = 2p_0\,\delta (\vec p - \vec p\,')\,\delta
_{mm'}\>,
\label{RIA}
\end{eqnarray}
Here $\vec p_1 \>,\>\vec p_2$ --- are particle momenta,
$m_1\>,\>m_2$ --- spin projections.

Since we consider the two-quark system as one composite system,
then the natural basis is one with separated center-of-mass
motion:
\begin{equation} |\,\vec P,\>\sqrt
{s},\>J,\>l,\>S,\>m_J\,>\>,
\label{bas-cm}
\end{equation}
with $P_\mu = (p_1 +p_2)_\mu$,
$P^2_\mu = s$, $\sqrt {s}$ ---
the invariant mass of two-particle system , $l$
--- the angular momentum in the center-of-mass frame,
$S$ --- total spin,
$J$ --- total angular momentum, $m_J$ ---
projection of total angular momentum.

The basis (\ref{bas-cm}) is connected with (\ref{RIA})
through the Clebsch -- Gordan decomposition of the Poincar\'e
group. The details of this procedure one can find in
\cite{KoT72}.
Now the decomposition of the electromagnetic current matrix
element for the composite system (\ref{pi-par})
in the basis (\ref{bas-cm}) has the form

$$
(p_c + p_c ')_\mu\,F_c (Q^2) = \sum \,\int \,\!\frac {d\vec P}
{N_{C-G}}
\,\frac {d{\vec P}'}
{N'_{C-G}}
\,d\sqrt {s}\,d\sqrt {s'}
<
p_c |\vec P,\sqrt {s},J,l,S,m_J>\cdot $$
\begin{equation}
<\vec P,\sqrt s,J,l,S,m_J \mid j_\mu \mid \vec P',\sqrt{s'}
,J',l',S',{m_J}'>\cdot
\label{pi-cur}
\end{equation}
$$
<\vec P',\sqrt{s'},J',l',S',{m_J}'|{p_c}'>.
$$
Here the sum is over the discrete variables of the basis
(\ref{bas-cm}).\\ $ <\vec P\,,\,\sqrt {s},\,J,\, l,\,S,\,m_J|p_c >$ -
is the composite system wave function
\begin{equation}
<\vec P\,',\,\sqrt
{s'},\,J',\, l',\,S',\,m_J'|\,p_c> =
{N_c}
\delta (\vec P\,' -
\vec p_c)\delta _{JJ'}\delta _{m_Jm_J'}
\delta _{ll'}\delta _{SS'}\,\varphi
^J_{lS}(k)\>.
\label{vf}
\end{equation}
$k = \sqrt{{(s^2-2s(M^2_{\bar s}+M^2_u)+\eta^2)}/{4s}}\>,
\quad \eta = M^2_q-M^2_{\bar Q}. \quad
N_c, N_{C-G}
$
are factors due to normalization. Concrete form of $N_c$ and $N_{C-G}$
will not be used.

Let us discuss the current operator matrix element which
enters the r.h.s. of the equation
(\ref{pi-cur}).

In the case of non-interacting quark system electromagnetic current
matrix element of this system can be parametrized similarly to the standard
case of one-particle matrix element, e.g. for
meson (\ref{pi-par}),
i.e. it is possible to extract the invariant part -- form factor $g_0$:
\begin{eqnarray}
<\vec P,\sqrt s,J,l,S,m_J |\,j^0_\mu\,| \vec P',\sqrt{s'}
,J',l',S',{m_J}'>=\nonumber\\ = A_\mu (s,Q^2,s')\> g_0(s,Q^2,s').
\label{param}
\end{eqnarray}

The vector $A_\mu (s,Q^2,s')$
is defined by the current transformation properties
(by the Lorentz--covariance and the current conservation law):

\begin{equation}
A_\mu =(1/Q^2)[(s-s'+Q^2)P_\mu + (s'-s+Q^2) P\,'_\mu
].  \label{vec-a}
\end{equation}

In our parametrization the
current is conserved by construction:
\begin{equation}
A_\mu(s,Q^2,s') Q^{\mu} = 0.
\label{conserv}
\end{equation}

In the frame of basis (\ref{RIA})
non-interacting current matrix element has the following form:
\begin{eqnarray}
<\vec p_1,m_1;\vec
p_2,m_2
|j^0_\mu|\vec p\,'_1,m'_1;\vec p\,'_2,m'_2>=\nonumber\\
=<\vec p_1,m_1|\vec p\,'_1,m'_1><\vec p_2,m_2|j_\mu|\vec p\,'_2,m'_2>+
  (1\leftrightarrow2).
\label{melRIA}
\end{eqnarray}
This is, as a matter of fact, the relativistic
impulse approximation. The one-particle current in (\ref{melRIA}) is
expressed in terms of one-quark form factors. Clebsh-Gordan
decomposition of the basis (\ref{bas-cm}) into basis (\ref{RIA}) gives the
expression of free form factor $g_0(s,Q^2,s')$ in terms of one-quark form
factors:
$$g_0(s,Q^2,s')=\frac{\sqrt{ss'}}{\sqrt{[s^2-2s(M_{\bar s}^2+M_u^2)+\eta^2]
[s'^2-2s'(M_{\bar s}^2+M_u^2)+\eta^2]}}\cdot $$
\begin{equation}
\label{ff-noint}
\cdot \frac{Q^2(s+s'+Q^2)}{2[\lambda(s,-Q^2,s')]^{3/2}} \cdot
\left (B^u(s,Q^2,s') + B^{\bar s}(s,Q^2,s')\right )\>,
\end{equation}
$$B^{\bar s}(s,Q^2,s') = \left [ f_1^{(\bar s)}(s+s'+Q^2-2\eta)\cos(\omega_1
+ \omega_2)-\right.$$
$$\left. -f_2^{(\bar s)}\frac{M_{\bar s}}{2}\xi (s,Q^2,s')
\sin(\omega_1 +\omega_2 )\right ]\,\theta (s,Q^2,s')\>,$$
$$
\xi (s,Q^2,s') = \sqrt{-\lambda (s,-Q^2,s')M_{\bar s}^2+ss'Q^2-\eta Q^2(s+s'+
Q^2)+Q^2\eta^2}\>,$$
$$\lambda(a,b,c)=a^2+b^2+c^2-2(ab+ac+bc)\>,
$$
$$f_1^{(\bar s)}=\frac{2M_{\bar s}\,G_E^{(\bar s)}(Q^2)}{\sqrt{4M_{\bar
s}^2+Q^2}};\qquad
f_2^{(\bar s)}=-\frac{4\,G_M^{(\bar s)}(Q^2)}{M_{\bar s}\sqrt{4M_{\bar
s}^2+Q^2}}\>,$$
$$\omega_1=\hbox{arctg} \frac{\xi (\,s\,,Q^2\,,s')
}{M_u[(\sqrt{s}+\sqrt{s'})^2+Q^2]
+(\sqrt{s}+\sqrt{s'})(\sqrt{ss'}+\eta)}\>,$$
$$\omega_2=\hbox{arctg} [(\sqrt{s}+\sqrt{s'}+2M_{\bar s})\,
\xi (\,s\,,Q^2\,,s')\cdot$$
$$\{M_{\bar s}(s+s'+Q^2)(\sqrt{s}+\sqrt{s'}+2M_{\bar s})+\sqrt{ss'}
(4M^2_{\bar s}+Q^2)
-\eta [2M_{\bar s}(\sqrt{s}+\sqrt{s'})-Q^2]\}^{-1}]\>,$$
$$\theta (s,Q^2,s') = \vartheta (s'- s_1) - \vartheta (s'- s_2)\>,$$
Here $\vartheta$ is the standard step function, $G_E^{(\bar s)}(Q^2)$
and $G_M^{(\bar s)}(Q^2)$ -- Sachs form factors, $\omega_1$ and
$\omega_2$ -- are the Wigner rotation parameters.
$$s_{1,2}=M_{\bar s}^2+M_u^2+\frac{1}{2M_{\bar s}^2}(2M_{\bar
s}^2+Q^2)(s-M_{\bar s}^2-M_u^2) \mp $$
$$\mp\frac{1}{2M_{\bar s}^2}\sqrt{Q^2(4M_{\bar s}^2+Q^2)[s^2-2s(M_{\bar
s}^2+M_u^2)+\eta^2]}\>.$$
Function $B^u(s,Q^2,s')$ can be deduced from $B^{\bar s}(s,Q^2,s')$ by
substitution $M_{\bar s} \leftrightarrow M_u.$

Let us return now to the Eq.(\ref{pi-cur}). The current matrix element
entering the r.h.s. of Eq.(\ref{pi-cur}) must be interaction dependent.
This dependence is known \cite{ChC88}, \cite{CoO75}
to be a consequence of the
current conservation law and of the condition of current
relativistic covariance. This means that we can not use
in Eq. (\ref{pi-cur}) the parametrization of non-interacting current
matrix element (\ref{param}) directly and need to include the interaction.
Let us perform the interaction including in (\ref{param}) in minimal
manner:  we shall include the interaction only in the vector function
$A_\mu (s,Q^2,s')$ in Eq.(\ref{vec-a}):
$$ A_\mu (s,Q^2,s')\>\to\> \frac{N_{C-G} N'_{C-G}} {N_cN'_c} A_\mu
^{int}$$
$$ A_\mu ^{int}  = A_\mu (s,Q^2,s')\left| _{_{P_\mu \to p_{c \mu},\,\,
P'_\mu \to p'_{c \mu}}} = (p'_c + p_c)_\mu \right.\>, $$ \begin{equation}
g_0(s,Q^2,s')\>\to\>g(s,Q^2,s') =
g_0(s,Q^2,s')\>.\label{a-int}
\end{equation}

The function $A_\mu ^{int}$
contains the interaction through the impulses $p'_{c \mu}$ and $p_{c
\mu}$.  Using (\ref{pi-cur}), (\ref{param}) and (\ref{a-int}) we obtain
now the following expression for the form factor:
\begin{equation}
F_c (Q^2)=\int d\sqrt s\ d\sqrt{s'}\ \varphi (k)\,g_0(s,Q^2,s')\,
\varphi (k').
\label{ff}
\end{equation}
Here we use for simplicity the notation:
$ \varphi ^J_{lS}(k)\> \to \> \varphi (k). $

Let us emphasize
, that
the r.h.s. of Eq.(\ref{pi-cur}) with (\ref{a-int}) inserted
satisfies the current conservation law: it is orthogonal to the
vector ${Q_\mu = (p'_c - p_c)_\mu}$. This latter fact is rather
noticeable because generally the construction of the
conserved current operator for composite systems presents
a rather complicated problem which is not solved yet
\cite{Lev95}, \cite{GrH92}.
Thus, the Eq.(\ref{ff}) takes into account the relativistic
covariance and the current conservation law. This is right for
any choice of the function $g(s,\,Q^2,\,s')$,
including the expressions (\ref{ff-noint}), (\ref{a-int}) which we use
here.

For $\varphi (k)$ one can use  any phenomenological
wave function, normalized using the relativistic density of
states:
$\varphi(k) =\sqrt{\sqrt{s}(1 - \eta^2/s^2)}\,u(k)\,k,$
$u(k)$ - is nonrelativistic phenomenological wave function.

Let us discuss the numerical results.

The Eq. (\ref{ff}) was used to calculate pion and kaon form factors. The
following wave functions were utilized:

1. A gaussian or harmonic oscillator (HO)
wave function (see e.g.  \cite{ChC88})
\begin{equation}
u(k)= N_{HO}\,
\hbox{exp}\left(-{k^2}/{2b^2}\right).
\label{HO-wf}
\end{equation}

2. A power-law (PL) wave function (see e.g. \cite{Sch94b},
\cite{CaG94}) \begin{equation}
u(k) =N_{PL}\,{(k^2/b^2 +
1)^{-n}}\>,\quad n = 2\>,3\>.  \label{PL-wf}
\end{equation}

3. The wave function with linear confinement from Ref.\cite{Tez91}:
\begin{equation}
u(r) = N_T \,\exp(-\alpha r^{3/2} - \beta r)\>,\quad \alpha =
\frac{2}{3}\sqrt{2\,M_r\,a}\>,\quad \beta = M_r\,b\>.
\label{Tez91-wf}
\end{equation}
$a\>,b$ -- parameters of linear and Coulomb parts of potential
respectively, $M_r$ -- reduced mass of two-particle systems.

Our point of view is that the choice of parameters is to be done in such a
way as to give the experimental value of MSR, because MSR can be determined
model independently by direct experiment on elastic meson-electron
scattering.
Using the definition of MSR  we obtain the condition:
\begin{equation}
<r^2>=\left.-\,6\,dF_\pi(Q^2)/dQ^2 \right|_{Q^2=0} = <r^2>_{exp}
\label{rad}
\end{equation}
so that the parameters in question are not more independent \cite{KrT93}.
They are connected through Eq. (\ref{rad}) which must be fulfilled
within
the experimental errors of r.h.s.

In our calculations we have fixed the masses of u- and d-quarks to be
equal to 0.25 GeV, and that of s-quark from the approximate relation
$M_s/M_u \approx$ 1.4. These values are usually used in relativistic
calculations.  Once the masses are fixed the Eq. (\ref{rad}) fixes
parameters $b$ in the models (\ref{HO-wf}) and (\ref{PL-wf}) or parameter
$a$ in the model (\ref{Tez91-wf}). We have used $b = $ 0.7867 for the
model (\ref{Tez91-wf}) which means that the strong coupling constant is
equal to 0.59 on the light meson mass scale. We have supposed the quark
anomalous
 magnetic moment to be zero.

The results for pion form factor at large $Q^2$ are given on fig.1. If MSR
is accurately given,
then one can see that the results depend strongly on the
choice of model. This dependence is much more pronounced than the
dependence on the parameters variation in the frame of one chosen model
within the experimental error of MSR in (\ref{rad}). This means that the
experiments on pion form factor at large $Q^2$ planned to be realized on
CEBAF in fact will enable one to choose between different two-quark pion
models.

The situation with kaon form factor is quite different. This is due to the
fact that kaon MSR is measured with rather less accuracy than that of pion
($\sim$ 15\% and $\sim$ 2\% respectively ) and then the Eq.(\ref{rad}) in
the kaon case does not give such strong constraints on the
parameters as in the case of pion. So the variations of the values of kaon
form factor at large $Q^2$ compatible with (\ref{rad}) are large.
One can see from the fig. 2 that the model dependence of kaon form factor
is strong at large $Q^2$, however for the model discrimination one needs
MSR and kaon form factor at low-momentum transfers to be measured with much
greater accuracy ($\sim$~2\% as in the case pion). This
problem  can not be ignored in other approaches too.

It is possible to make the following conclusions:

1. The relativistic approach used in this Letter gives pion and kaon form
factors in good agreement with the available data.
Relativistic effects are
large at all values of momentum transfers.

2. The pion and kaon form factors at large $Q^2$ depend strongly on the
choice of model. The experiments on pion form factor at large momentum
transfer planned at CEBAF will choose between such models.

3. In the case of kaon such a choosing may be performed only if
supplemented  by accurate measurements of kaon MSR.\\[2mm]
This work is supported in part by State Committee for Higher
Education of Russia
grant no.94-6.7-2015.

\newpage
\pagestyle{empty}
\begin{center}
{\bf Figure Captions}
\end{center}

Fig. 1. Electromagnetic pion form factor, $Q^2 F_{\pi}(Q^2)$,
at high momentum transfer
in different models ($M_u = M_d = 0.25$ $GeV/c$). \\
1 -- harmonic
oscillator wave function Eq.(\ref{HO-wf}), $b= $ 0.207 GeV;  \\
2 -- power-law
wave function Eq.(\ref{PL-wf}), $n = $ 2,  $b= $ 0.274 GeV; \\
3 -- power-law wave
function Eq.(\ref{PL-wf}), $n = $ 3,  $b= $ 0.388 GeV; \\
4 -- wave function Eq.(\ref{Tez91-wf}) from
model with linear confinement \cite{Tez91},\\  $a =$
0.0183 $\hbox{GeV}^2$, $b= $  0.7867.\\
Experimental data are taken from Ref. \cite{Ame84}, \cite{Beb78}.
\medskip

Fig. 2. Electromagnetic kaon form factor at high momentum transfer  in
different models.
The same line code as in Fig.1 is used.
$M_s =$ 0.35
GeV, 1 -- $b = $ 0.255 GeV, 2 -- $b =$ 0.339 GeV, 3 -- $b =$ 0.480 GeV,
4 -- $a= $ 0.0318 $\hbox{GeV}^2$, $b= $  0.7867.\\
Experimental data are taken from Ref. \cite{Ame86}

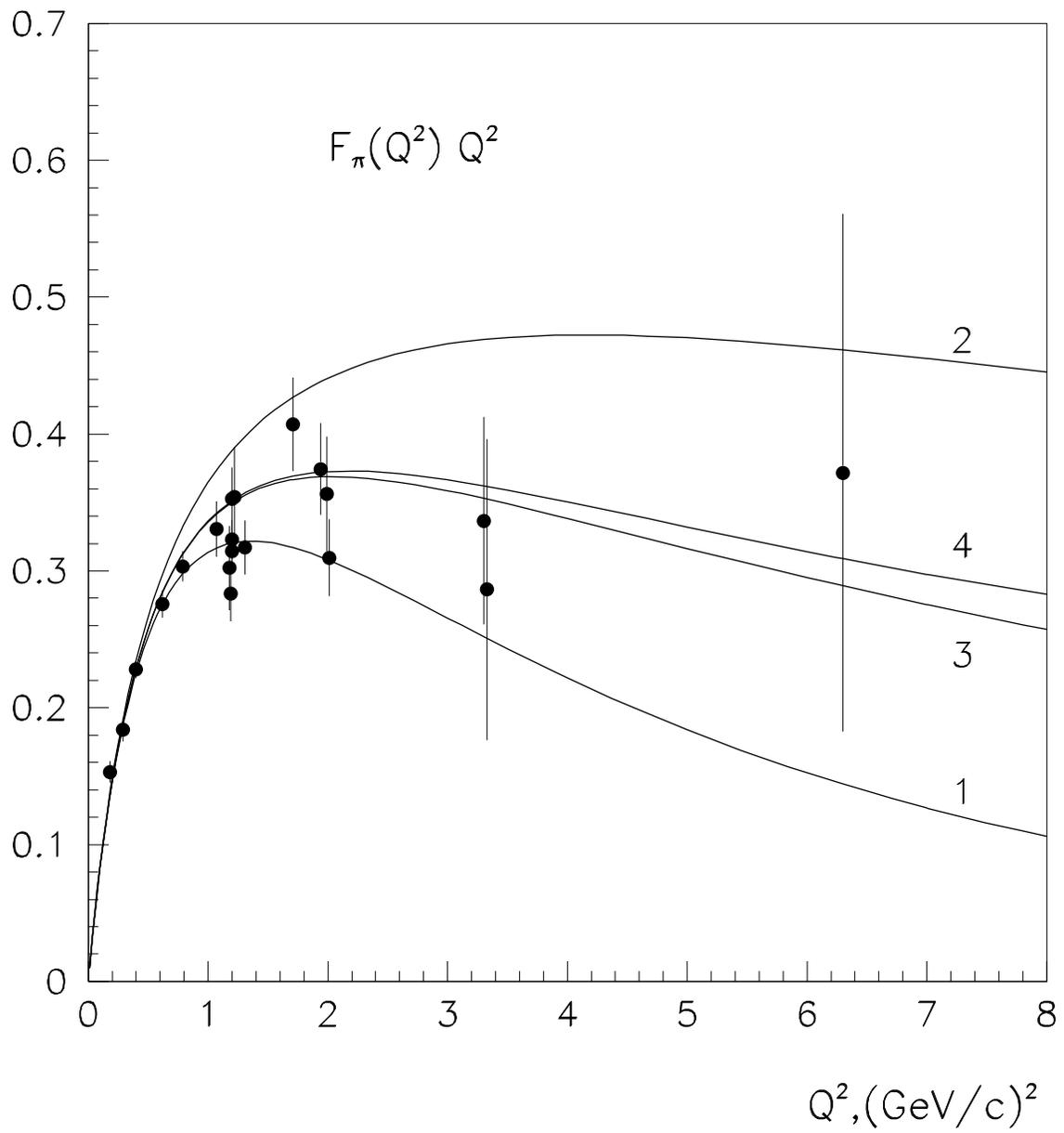
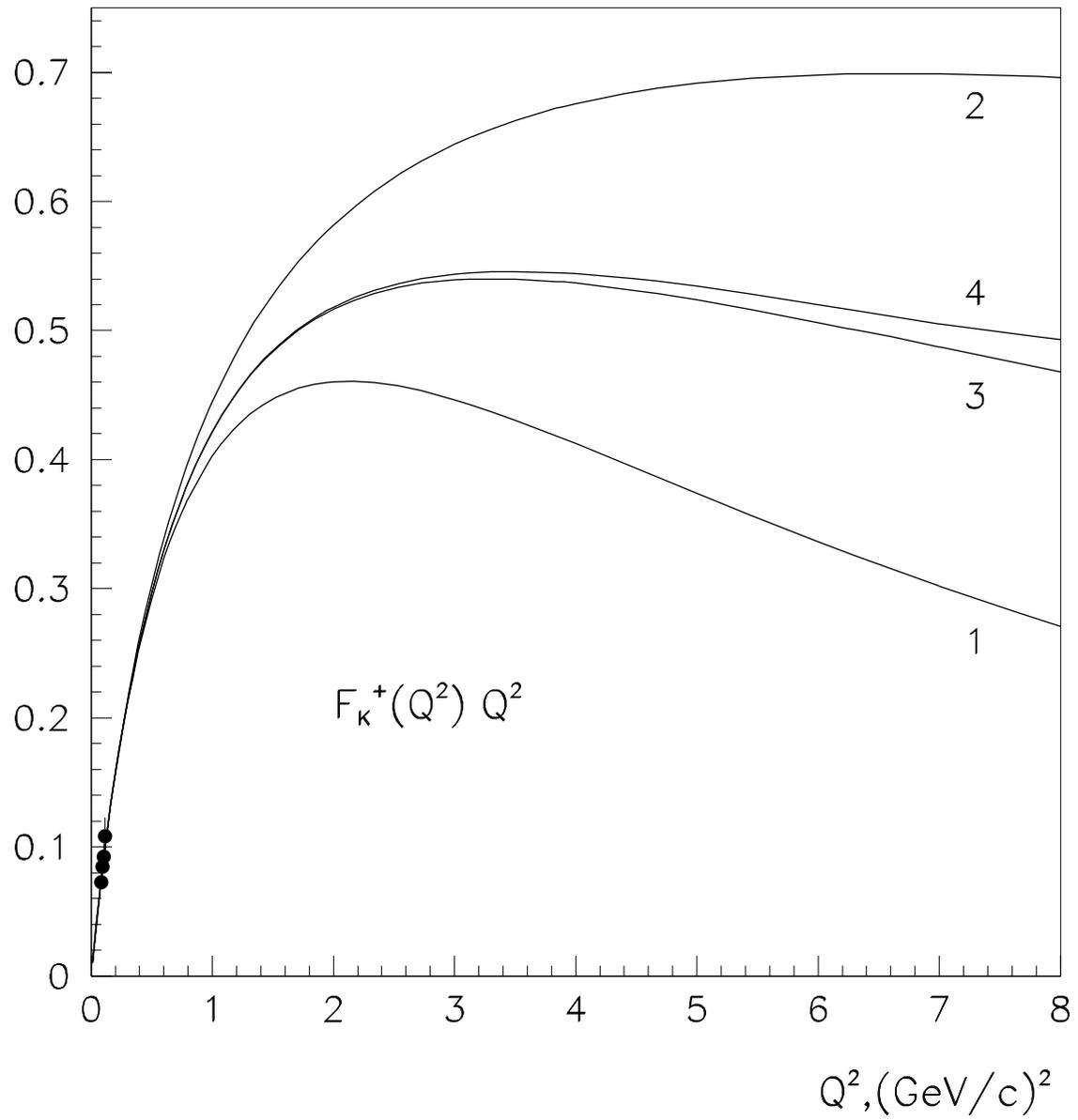
\end{document}